\documentstyle[12pt,aaspp4]{article}
\newcommand{\kteffemerge}{\kteff}
\makeatletter

\@ifundefined{chapter}{\def\thebibliography#1{\section*{References\@mkboth
  {REFERENCES}{REFERENCES}}\list
  {\relax}{\setlength{\labelsep}{0em}
        \setlength{\itemindent}{-\bibhang}
        \setlength{\itemsep}{0pt}
        \setlength{\parsep}{0pt}
        \setlength{\leftmargin}{\bibhang}}
    \def\newblock{\hskip .11em plus .33em minus .07em}
    \sloppy\clubpenalty4000\widowpenalty4000
    \sfcode`\.=1000\relax}}%
{\def\thebibliography#1{\chapter*{Bibliography\@mkboth
  {BIBLIOGRAPHY}{BIBLIOGRAPHY}}\list
  {\relax}{\setlength{\labelsep}{0em}
        \setlength{\itemindent}{-\bibhang}
        \setlength{\itemsep}{0pt}
        \setlength{\parsep}{0pt}
        \setlength{\leftmargin}{\bibhang}}
    \def\newblock{\hskip .11em plus .33em minus .07em}
    \sloppy\clubpenalty4000\widowpenalty4000
    \sfcode`\.=1000\relax}}

\newlength{\bibhang}
\let\@internalcite\cite
\def\cite{\let\@citeleft(\let\@citeright)%
    \@ifstar{\citeyear}{\citefull}}
\def\acite{\let\@citeleft\relax\let\@citeright\relax%
    \@ifstar{\citeyear}{\acitefull}}
\def\citenp{\let\@citeleft\relax\let\@citeright\relax
    \@ifstar{\citeyear}{\citefull}}
\def\citefull{\def\astroncite##1##2{##1~##2}\@internalcite}
\def\citeyear{\def\astroncite##1##2{##2}\@internalcite}
\def\acitefull{\def\astroncite##1##2{##1~(##2)}\@internalcite}

\def\@citex[#1]#2{\if@filesw\immediate\write\@auxout{\string\citation{#2}}\fi
  \def\@citea{}\@cite{\@for\@citeb:=#2\do
    {\@citea\def\@citea{; }\@ifundefined
       {b@\@citeb}{{\bf ?}\@warning
       {Citation `\@citeb' on page \thepage \space undefined}}%
{\csname b@\@citeb\endcsname}}}{#1}}

\def\@cite#1#2{\@citeleft#1\if@tempswa , #2\fi\@citeright}
\def\@biblabel#1{}

\makeatother

\newcommand{\PSbox}[3]{\mbox{\rule{0in}{#3}\includegraphics{#1}\hspace{#2}}}
\newcommand{\FigNum}[1]{\unitlength 1pt \begin{picture}(55,10)(-400,35) 
                        \put(0,0){Figure #1}
                        \end{picture}}
\newcommand{\msun}{$M_\odot$} 
\newcommand{\zsun}{$Z_\odot$} 
\newcommand{\persec}{\mbox{$\second^{-1}$}}
\newcommand{\percm}{\mbox{$\cm^{-2}$}}
\newcommand{\ppm}{\mbox{$\pm$}}
\newcommand{\cgsflux}{\erg\percm\persec}

\newcommand\approxlt{\mbox{$^{<}\hspace{-0.24cm}_{\sim}$}}
\def\etal{{et~al.}}

\newcommand{\nh}{\mbox{$N_{\rm H}$}}
\newcommand{\nhtt}{\mbox{$N_{\rm H, 22}$}}
\newcommand{\ud}[2]{\mbox{$^{+ #1}_{- #2}$}}
\newcommand{\ee}[1]{\mbox{$10^{#1}$}}
\newcommand{\tee}[1]{\mbox{$\times 10^{#1}$}}

\newcommand{\perval}[2]{{#1\mbox{$^{#2}$}}} 

\def\chisqrnu{\mbox{$\chi^2_\nu$}}
\def\x1608{{4U~1608$-$522}}

\def\cenx4{{Cen~X$-$4}}
\def\aql{{Aql~X$-$1}}
\def\saxj1808{{SAX J1808.4$-$3658}}

\newcommand{\cm}{\mbox{$\rm\,cm$}}

\newcommand{\second}{\mbox{$\rm\,s$}}

\newcommand{\erg}{\mbox{$\rm\,erg$}}

\newcommand{\kpc}{\mbox{$\rm\,kpc$}}

\newcommand{\kteff}{\mbox{$kT_{\rm eff}$}}
\newcommand{\kteffinfty}{$kT_{\rm eff, \infty}$}

\newcommand{\rinfty}{\mbox{$R_{\infty}$}}
\newcommand{\rproper}{\mbox{$R$}}

\newcommand{\chandra}{{\em Chandra\/}}

\newcommand{\rosat}{{\em ROSAT\/}}
\newcommand{\asca}{{\em ASCA\/}}

\newcommand{\beppo}{{\em BeppoSAX\/}}
\newcommand{\sax}{\beppo}

\begin{document}

\title{Quiescent Thermal Emission from the Neutron Star in Aql X-1}

\author{Robert E. Rutledge\altaffilmark{1}, 
Lars Bildsten\altaffilmark{2}, Edward F. Brown\altaffilmark{3}, 
George G. Pavlov\altaffilmark{4}, 
\\ and Vyacheslav  E. Zavlin\altaffilmark{5}}
\altaffiltext{1}{
Space Radiation Laboratory, California Institute of Technology, MS 220-47, Pasadena, CA 91125;
rutledge@srl.caltech.edu}
\altaffiltext{2}{
Institute for Theoretical Physics and Department of Physics, Kohn Hall, University of 
California, Santa Barbara, CA 93106; bildsten@itp.ucsb.edu}
\altaffiltext{3}{
Enrico Fermi Institute, 
University of Chicago, 
5640 South Ellis Ave, Chicago, IL  60637; 
brown@flash.uchicago.edu}
\altaffiltext{4}{
The Pennsylvania State University, 525 Davey Lab, University Park, PA
16802; pavlov@astro.psu.edu}
\altaffiltext{5}{ 
Max-Planck-Institut f\"ur Extraterrestrische Physik, D-85740 Garching,
Germany; zavlin@xray.mpe.mpg.de}

\begin{abstract}

We report on the quiescent spectrum measured with \chandra/ACIS-S of
the transient, type-I X-ray bursting neutron star \aql, immediately
following an accretion outburst. The neutron star radius, assuming a
pure hydrogen atmosphere and hard power-law spectrum, is
\rinfty=13.4\ud{5}{4} $(d/5 \kpc)$ km.  Based on the historical
outburst record of {\em RXTE}/ASM, the quiescent luminosity is
consistent with that predicted by Brown, Bildsten and Rutledge from
deep crustal heating, lending support to this theory for providing a
minimum quiescent luminosity of transient neutron stars.  While not
required by the data, the hard power-law component can account for
18\ppm8\% of the 0.5-10 keV thermal flux.  Short-timescale intensity
variability during this observation is less than 15\% rms (3$\sigma$;
0.0001-1 Hz, 0.2-8 keV). Comparison between the \chandra\ spectrum and
three X-ray spectral observations made between Oct 1992 and Oct 1996
find all spectra consistent with a pure H atmosphere, but with
temperatures ranging from 145--168 eV, spanning a factor of
1.87\ppm0.21 in observed flux. The source of variability in the
quiescent luminosity on long timescales (greater than years) remains a
puzzle. If from accretion, then it remains to be explained why the
quiescent accretion rate provides a luminosity so nearly equal to that
from deep crustal heating.

\end{abstract}


\section{Introduction}

  Brown, Bildsten \& Rutledge \cite*[BBR98 hereafter]{brown98} showed
that the core of a transiently accreting neutron star (NS), such as
Aql X-1 (for reviews of transient neutron stars, see
\citenp{chen97,campana98b}), is heated by nuclear reactions deep in
the crust during the accretion outbursts. The core is heated to a
steady-state in $\sim$\ee{4} yr (see also \citenp{colpi00}), after
which the NS emits a thermal luminosity in quiescence of
(BBR98)
\begin{equation}
\label{eq:brown}
L_q = 8.7\times10^{33} \left(\frac{\langle \dot{M} \rangle}{10^{-10}
M_\odot {\rm yr}^{-1}}\right) \frac{Q}{1.45 {\rm MeV}/m_p} \; \; {\rm ergs \; s}^{-1},
\end{equation}
where $\langle \dot{M} \rangle$ is the time-averaged mass-accretion
rate onto the NS, and $Q$ is the amount of heat deposited in the crust
per accreted nucleon  (\citenp{haensel90}; see
\citenp{bildstenrutledge00} for a discussion).

 Aql X-1 has been detected in X-ray quiescence five times: once with
the \rosat/HRI and twice with \rosat/PSPC \cite{verbunt94}, once with
\asca\ \cite{asai98}, and once with \sax\ \cite[C98
hereafter]{campana98a}.  We report in this {\em Letter} the detection
and CCD X-ray spectroscopy of a \chandra\ observation of \aql\ in
quiescence, and compare its luminosity and spectrum to three of these
previous observations.  We show that the quiescent X-ray spectrum is
consistent with thermal emission from a pure H atmosphere on the NS
\cite{rajagopal96,zavlin96}, as is observed from this and other
transient neutron stars in quiescence (Rutledge \etal\
\citenp*{rutledge99,rutledge00,rutledge01}). As pointed out by BBR98,
for accretion rates \approxlt 2\tee{-13} \msun\perval{yr}{-1}, gravity
stratifies metals in the NS atmosphere faster than they can be
provided by accretion (Bildsten, Salpeter, \& Wasserman
\citenp*{bildsten92}), making a pure H atmosphere the appropriate
description of the NS photosphere.

  Callanan, Filippenko \& Garcia \cite*{callanan99} recently
showed that the optical counterpart to Aql X-1 is a faint star
near the previously mis-identified counterpart. This led to the
counterpart's identification as a late type star (spectral type K7 to
M0) with a quiescent magnitude $V=21.6$ at a reddening of $A_V\approx
1.6$ \cite{chev99}. The orbital period has been well measured at
$P_{\rm orb}=18.95 \ {\rm hr}$ via photometric observations both in
outburst \cite{chevalier98,garcia99b} and quiescence \cite{welsh00}.

  Chevalier et al. \cite*{chev99} estimated the distance to the binary
as 2.5 kpc by assuming that the counterpart was a main sequence star
of spectral type K7, or roughly $M_c\approx 0.6 M_\odot$ and
$R_c\approx 0.6 R_\odot$. However, the steady mass transfer and
ellipsoidal variations \cite{welsh00} require that the companion be
Roche-lobe filling, which gives $R_c\approx
1.65R_\odot(M_c/M_\odot)^{1/3}$, larger than Chevalier et al.'s
estimate. If we use the de-reddened quiescent $V$ magnitude, we find
that the distance to the binary is $d=4.7(6.4) \ {\rm kpc}$ (for
spectral type K7) for $M_c=0.5(1.0)M_\odot$. For a spectral type of
M0, we find $d=4(5) \ {\rm kpc}$ for $M_c=0.5(1.0)M_\odot$.  Thus, the
clear minimum distance is 4 kpc, and the current uncertainties allow
for a distance as large as 6.5 kpc. The Type I bursts observed by
Czerny, Czerny and Grindlay \cite*{czerny87} had peak fluxes of
$\approx 7\times 10^{-8} \ {\rm erg \ cm^{-2} \ s^{-1}}$, which gives
the solar abundance Eddington luminosity of $2\times 10^{38} \ {\rm
erg \ s^{-1}}$ at $d=5 \ {\rm kpc}$.  We use 5 kpc as our fiducial
distance.

In \S~\ref{sec:anal}, we describe the \chandra\ observation and the
constraints on variability during the observation. Section
\ref{sec:spectral} shows our results on the spectral analysis and
compares the \chandra\ observations to previous quiescent
observations. We conclude \S \ref{sec:spectral} with a comparison of
the observed luminosity to that predicted by equation
(\ref{eq:brown}). We discuss these results in the context of alternate
emission mechanisms in \S~\ref{sec:con}.

\section{Chandra Observations and Timing Analysis}
\label{sec:anal}

\aql\ was observed with \chandra\ \cite{chandra} using the ACIS-S3
detector in imaging mode, beginning Nov 28 2000, 10:51:35 UT, 7 days
after the last 1-day detection (3$\sigma$, $\approx$20 mCrab) with
RXTE/ASM \cite{asm} during the outburst.

The X-ray source position was offset 4\arcmin\ from the optical-axis
to mitigate pileup. The observation had a total exposure time of
6627.6 s over a period of 7307.6 sec (90\% livetime), with time
resolution of 0.44104 sec, and a 0.4 sec exposure.  One X-ray source
is detected in the field, at a position (based on spacecraft pointing)
$\alpha=$ 19h11m16.00s, $\delta=$00d35m06.4s (J2000), with systematic
errors dominating the uncertainty of position (\ppm 1\arcsec),
consistent with the known optical position of \aql\
\cite{thorstensen78,callanan99}.

The data were analyzed using the CIAO v2.0
\footnote{http://asc.harvard.edu/ciao2.0/} and XSPEC v11 \cite{xspec}.
X-ray source counts were extracted within an area 10 pixels in radius
about the source position, with a total of 1243 counts. At 0.075
counts/frame, the pileup fraction is $<$3\% and can be neglected.
Background was taken from an annulus centered on the source position,
with radii of 13 and 50 pixels. The expected number of background
counts in the source region is 8 counts, which are neglected in our
analyses.

We binned the 0.2-8 keV counts into three light curves with three
different time resolutions: $\delta T$=0.44104 s, 10 s, and 100 s.  In
each case, the distribution of the number of counts per bin was
consistent with a Poissonian distribution, for an average number of
counts per bin of (1237 counts)/(7308 s/$\delta T$).
We also produced a power density spectrum (PDS), beginning with a
lightcurve of all the counts in the source region with time resolution
of 0.44104 sec, producing a 0.00014-1.13 Hz PDS.  The PDS is
statistically consistent with a constant power at the Poisson
level, and shows no evidence of any excess variability.   Fitting
the PDS with a power-law component with a fixed slope ($\propto
\nu^{-1}$) above a (fixed) Poisson level gives a 3$\sigma$ upper-limit
to the root-mean-square variability of $<$15\% (0.0001-1 Hz; 0.2-8
keV). We thus find no evidence of  intensity
variability during the \chandra\ observation.

\section{Spectral Analysis}
\label{sec:spectral}

We binned the data into 13 energy bins (0.5-8.0 keV), and fit several
single component spectral models (powerlaw, H atmosphere, 
Raymond-Smith, multicolor disk or blackbody).  We also include H
atmosphere with a power-law, with two fixed values for the power-law
slope ($\alpha=1$, $\alpha=2$).  Galactic absorption (\nh=\ee{22}\nhtt
\perval{cm}{-2}) is initially left as a free parameter. The best-fit
models were all statistically acceptable, and most gave column
densities consistent with $A_{\rm V}=1.6$, and \nhtt=0.179 $A_{\rm V}$
\cite{gorenstein75,predehl95}; the parameters are given in
Table~\ref{tab:fits}.

The Raymond-Smith spectrum has a metallicity substantially below the
solar value ($Z<$5\tee{-3} \zsun). In addition, Bildsten and Rutledge
(2000) showed that the quiescent X-ray-to-optical flux ratio of \aql\
($\sim$\ee{0.5\ppm0.2}) is much greater than the maximum observed from
quiescent stellar coronae ($F_X/F_{\rm opt}$\approxlt\ee{-3}), so we
exclude the stellar coronae as a possible solution. The single
power-law spectrum requires a higher column density than the optically
implied value (\nhtt=0.72 vs. 0.30) and is much steeper (photon slope
$\alpha$=4.1) than is typically observed from non-thermal X-ray
sources.  We reject the model on this basis.

Standard Shakura-Sunyaev \cite{shakura73} disks have largely been
excluded as the dominant emission of transients in quiescence
\cite{mcclintock95}, although recent interest in alternative disk
models \cite{nayakshin01} suggest it is useful to constrain parameters
for a multicolor disk model.  And, although black-body emission does
not physically describe the emergent spectrum from a transient neutron
star atmosphere in quiescence (BBR98), we include these spectral
parameters as well for the interested reader. Neither $R_{\rm in}$ of
the multicolor disk model nor \rinfty\ of the blackbody model appear
to correspond to any physically interesting values; both are smaller
than canonical NS radii. 

The simplest interpretation of the spectrum is thermal emission from a
pure Hydrogen atmosphere on the NS, and so we include the best-fit
parameters in Table~\ref{tab:fits}.  However, previous observations
have found a power-law component which dominates the quiescent
spectrum at high energies in \aql\ \cite{campana98a}, and in the
transient \cenx4\ \cite{asai96b,campana00,rutledge01}, with values
between $\alpha=1$ and 2.  We included a power-law component, with
slope fixed alternately at $\alpha=1$ and $\alpha=2$.  An F-test
\cite{press} shows that the additional spectral component does not
significantly improve the model fit (prob=0.07 and 0.13,
respectively).  The resulting values of \rinfty\ are systematically
higher, while \kteffinfty\ is systematically lower, indicating that
the presence of a weak power-law may be biasing the pure H atmosphere
spectral model parameters.  We therefore prefer the values from the H
atmosphere plus $\alpha=1$ spectral model (over the H atmosphere
spectral model alone),  of \rinfty=13.4\ud{5}{4} km and
\kteffinfty=135\ud{18}{12} eV.  These are not significantly different
from those of the $\alpha=2$ spectral model.

When we hold \nhtt\ fixed at its best-fit value, we obtain (for the
$\alpha=1$ spectrum) \rinfty=13.4\ppm2.0 km and
\kteffinfty=135\ud{10}{9} eV, indicating that the uncertainty in \nh\
has a modest effect on the uncertainty in the H atmosphere spectral
parameters.

Here \kteffinfty\ and \rinfty\ are the effective temperature and NS
radius as measured by a distant observer.  They obey the relation
$L_{\rm bol}^\infty=4\pi$\rinfty$^2\sigma T_{\rm eff, \infty}^4$,
where $L_{\rm bol}^\infty$ is the luminosity measured by a distant
observer.  The ``proper'' values of \kteffemerge\ and \rproper\ are
related to those at infinity through the gravitational redshift
parameter $g_{r} = [1-2GM/(Rc^{2})]^{1/2}$ (the redshift is $1+z =
g_{r}^{-1}$); for a neutron star of mass $1.4\,M_{\odot}$ and radius
$10\mathrm{\,km}$, $g_{r} = 0.766$.  The gravitational field redshifts
photon energies and bends their trajectories, so that
\kteffemerge=\kteffinfty$~g_r^{-1}$ and \rproper=\rinfty~$g_r$.  The
observed and proper bolometric luminosities are then related by
$L_{\rm bol}^\infty = L_{\rm bol} g_{r}^{2}$; one power of
$g_{r}$ accounts for the energy redshift and the other accounts for
the time dilation.  The observed bolometric thermal flux is then
$F_{\rm bol}^\infty$=2.5\tee{-12} \cgsflux.

\subsection{Comparison to Previously Measured Quiescent Spectra}

We simultaneously fit the \chandra\ spectrum with three energy spectra
taken in quiescence: \rosat/PSPC in October 1992, \rosat/PSPC in March
1993, and \asca\ in October 1996 (cf. Table~\ref{tab:obs}; see
\citenp{rutledge99} for details on these spectra).  We included a 4\%
systematic uncertainty, to allow for differing calibrations between
instruments.  The results of these spectral fits are in
Table~\ref{tab:results}).

Fitting an absorbed, pure H atmosphere spectrum with all identical
parameters produces a statistically unacceptable fit
(\chisqrnu/dof=2.63/67 dof; here and elsewhere in this paper, the
first number is the reduced $\chi^2$ value, the second number is the
number of degrees of freedom, and the two are separated by the
backslash character).  However, acceptable fits are found if we permit
either \nh (marginally), \kteffinfty\ or \rinfty\ to vary between the
four observations.  The observed 0.5-10 keV fluxes span a range of a
factor 1.87\ppm0.21 (90\% confidence).  Therefore, while the spectra
are statistically different, we are unable to observationally discern
which spectral parameter (or combination of parameters) is changing.

We then added the power-law component.  The best fit spectrum in which
all five spectral parameters are the same for the four different
observations was statistically unacceptable (\chisqrnu=2.80/65 dof;
prob=2\tee{-13}), indicating spectral variability. Acceptable fits are
found if we let any of the four parameters (\nhtt, \kteffinfty,
\rinfty, $F_{X, PL}$) vary independently between the four
observations.  The constraints on the power-law slope are very weak.
Also, in the fit in which $F_{X, PL}$ was permitted to vary, the value
$\alpha=3.5\ppm0.5$ is steeper than is typical of these sources,
combined with a high value of \nh.

We examined the temperature decrease as a function of time since the
outburst end.  Table~\ref{tab:obs} gives the time-delay between the
quiescent observation and the end of the previous outburst;
Table~\ref{tab:results} contains the temperatures measured during the
quiescent observation.  For the 1993 \rosat\ observation, we assumed
that the outburst ended 60 days after its start.  A linear fit to the
data finds a temperature which decreases ($-50$\ppm90) eV/yr
(consistent with no decrease), with a 3$\sigma$ upper-limit of $<$270
eV/yr. This is an uncertain measurement for a number of reasons: the
number of days since outburst start is uncertain by $\sim$7 days, the
duration of outbursts varies from outburst to outburst, and the
initial temperature of the NS atmosphere may be related to the total
fluence of the most recent outburst, which is unknown for 2 of the 4
outbursts. Moreover, the temperatures taken alone are not
significantly different. This level of decrease is consistent with the
limit on the decrease in the thermal temperature of \cenx4, which is
($-$2.2\ppm1.8) eV/yr \cite{rutledge01}.

\subsection{Power-law/Thermal Flux Ratio}
\label{sec:powerlaw}

C98 reported a power-law component that dominates the X-ray spectrum
at energies above $\approx$2 keV in quiescence, similar to a power-law
component seen in \cenx4\
\cite{asai96b,asai98,campana00,rutledge01}.  As we find
above, the \chandra\ spectrum does not require an additional power-law
component at high energies, although inclusion of a power-law
component makes a systematic difference to the parameters of the
thermal component.  We stress, therefore, that the power-law flux
values we discuss in this section are merely the best-fit values, and
are not evidence that the power-law component is present.

The best-fit power-law/thermal luminosity ratio is 18\ppm8\%
(26\ppm13\%) for a fixed $\alpha=1$ ($\alpha=2$; see
Table~\ref{tab:fits}).  The best-fit power-law flux is
(2.2\ppm1)\tee{-13} [(3.1\ud{1.6}{1.4})\tee{-13}] \cgsflux for
$\alpha=1$ [$\alpha=2$] in the 0.5-10 keV pass band. These are
consistent with power-law flux reported in quiescence from the
observations of C98 (5\tee{-13} \cgsflux).

We investigated if a variable power-law intensity could be responsible
for the variations between the four observations.  We held the value
of $\alpha$=1 fixed (the value found by C98 in observations 3-6),
kept the \nh, \kteffinfty, and \rinfty\ constant between the four
observations, and permit the power-law flux to vary between
observations.  The best fit was unacceptable (\chisqrnu/dof=1.64/65
dof; prob=8\tee{-4}).  We conclude that the observed spectral
variability cannot be explained by a variable power-law component with
the slope observed by C98.  The variability is caused either by a
power-law which changes in slope and flux, or variability in the
thermal component or column density.

\subsection{Predicted versus Observed Quiescent Flux}

\aql\ is one of the brightest known quiescent NSs, and, combined
with its frequent accretion outbursts, provides the best opportunity
to test the relationship between the time-averaged outburst flux and
the quiescent flux predicted by BBR98.

 We use the RXTE/ASM data to measure the time-averaged outburst flux
$\langle F \rangle $ (which is proportional to $\langle \dot M
\rangle$). We integrate all counts with $>5\sigma$ significance during
the five year period Jan 1996-Jan 2001 (3.9\tee{8} counts; we adopt a
\ppm10\% uncertainty), which includes $\approx$5 outbursts. If the
flux of Aql~X-1, when not in outburst, was always just below the
RXTE/ASM 1-day detection limit (0.1 c/s), this would increase the
time-average outburst flux by only 4\%. Using W3PIMMS, the ASM
counts/flux conversion for power-law spectra $\alpha$=0.8, 1.0, 2.0
are 3.4, 3.6, and 5.0\tee{-10} ergs\ \perval{cm}{-2}\perval{count}{-1}
(0.5-10.0 keV), respectively.  For blackbody spectra kT=0.8--1.4 keV,
the conversion factors are approximately the same: 3.3\tee{-10}
ergs\ \perval{cm}{-2}\perval{count}{-1}. We adopt 3.6\tee{-10} ergs\
\perval{cm}{-2}\perval{count}{-1} (0.5-10.0 keV) with the realization
that the spectral and bolometric uncertainties can be as large as
$\sim$25\%. This then gives $\langle F \rangle \approx 10^{-9}$
\cgsflux , or roughly $\langle \dot M \rangle \approx 2.5\times
10^{-10} M_\odot \ {\rm yr^{-1}}(d/5 {\rm kpc})^2$.

 We assume that the NS liberates $GM/R\approx 1.8\times 10^{20} \ {\rm
erg \ g^{-1}}$ per accreted baryon during the outburst. Then, from
equation (\ref{eq:brown}), we expect a quiescent bolometric thermal
flux $F_q\approx \langle F \rangle /130$, or $F_q=7.7\tee{-12}$
\cgsflux, compared with $F_{\rm bol}^\infty$=2.5\tee{-12} \cgsflux\
observed during the \chandra\ observation. The quiescent bolometric
thermal flux is uncertain by \ud{0.35}{0.26} dex (about a factor of 2
in both directions, 1$\sigma$), and so is consistent with this
prediction (BBR98; Eq. \ref{eq:brown}).

Note, on the other hand, that if one were to assume that 100\% of the
emergent luminosity were from accretion, this would set $\dot
M=1.1\tee{-12}$ \msun\perval{yr}{-1}, which is above the $\dot M$
where gravity stratifies metals in the NS atmosphere -- that is, where
metal lines would become unobservable.  We do not exclude the presence
of metal lines in the observed spectrum, and it is possible that, at
such an implied accretion rate, metal lines may be observed.

\section{Discussion and Conclusions}
\label{sec:con}

We have detected \aql\ in quiescence, at a luminosity comparable to
that observed in three previous epochs. 
This makes \aql\ the second transiently accreting, type-I
X-ray bursting NS (after \cenx4; \citenp{rutledge01}) which maintains a
quiescent luminosity to within a factor of a few.  The stability of
Cen X-4's quiescent luminosity over years, and the ability of Aql~X-1
to return to the same quiescent luminosity between outbursts (where
$\dot M$ increases by more than a factor of 1000) are both strong
support for their basal quiescent luminosity being set by deep crustal
heating (BBR98). In addition, spectral evidence points to most of the 
quiescent emission being thermal emission from the NS surface.

While the H atmosphere spectral model  implies a NS radius which
is at the low end of the EOS range for the adopted distance (see, for
example, \acite{lattimer01}), inclusion of an additional power-law
component -- although not statistically required -- produces a
systematically larger value of \rinfty\ and smaller value of
\kteffinfty, which is the expected direction of bias if the previously
observed power-law component is present.  We therefore quote the
best-fit value of \rinfty=13.4\ud{5}{4} (d/5 kpc) km and
\kteffinfty=135\ud{18}{12} eV which includes an $\alpha=1$ power-law
component, noting that these are not significantly different when we
assume a value $\alpha=2$.  Greater scrutiny of the distances to this
and similar objects is called for.  Moreover, we examined the spectrum
only in a limited range (0.5-8.0 keV), as \chandra/ACIS-S is not yet
calibrated down to 0.2 keV.

 However, puzzles remain. The quiescent X-ray luminosity of Aql X-1
has varied by a factor of 1.87\ppm0.21 over timescales of years, while
remaining constant ($<$15\% rms) on timescales 1-10,000 sec. This is
similar to the intensity variability observed from another quiescent
transient neutron star, \cenx4, which has shown variability of a
factor of $\approx$3 on timescales of days-yrs
\cite{campana97,rutledge00}, and $<$18\% rms variability on 1-10,000
sec timescales \cite{rutledge00}. In addition, the power-law spectral
components in some of these sources cannot be explained as thermal
emission; their origin is unclear.

A variation in \kteffinfty\ can be attributed to either a variation in
quiescent accretion onto the compact object or a changing thermal
emission.  Since the NS core cannot change its temperature on these
timescales, any variation in the thermal emission would need to be
from either internal heat sources previously neglected
\cite{ushomirsky01} or changes in the overlying envelope.

If quiescent accretion powers the variability in Cen X-4 and Aql X-1,
then we are left with the surprising coincidence that $\dot M_q$
provides a luminosity comparable to the deep crustal heating
luminosity (Eq.~1), in at least two different systems.  Even in its
simplest form (a diminished accretion rate from a cool disk,
\citenp{king98}) it seems surprising that quiescent accretion would
maintain near equality with the deep crustal heating luminosity. It
seems even less likely if one invokes a magnetic propeller to regulate
the accretion rate onto the NS in quiescence \cite{menou99}.
Therefore, if the variability is due to accretion, it remains to be
explained why the accretion rate provides a luminosity so nearly equal
to that from deep crustal heating.

\acknowledgements

The authors are grateful to the \chandra\ Observatory team for
producing this exquisite observatory.  We thank the anonymous referee
for comments which improved this paper.  This research was partially
supported by the National Science Foundation under Grant
No. PHY99-07949 and by NASA through grant NAG 5-8658, NAG 5-7017 and
the \chandra\ Guest Observer program through grant NAS GO0-1112B.
L. B. is a Cottrell Scholar of the Research Corporation.
E. F. B. acknowledges support from an Enrico Fermi Fellowship.


\clearpage
\pagestyle{empty}
\begin{figure}[htb]
\caption{ \label{fig:chandraspec} The $\nu F_\nu$ model spectrum of
\aql, and the observed \chandra/ACIS-S BI data.  The solid line is the
best-fit {\em unabsorbed} (the intrinsic X-ray spectrum of \aql, prior
to absorption by the interstellar medium; see Table~\ref{tab:fits})
H-atmosphere plus power-law model spectrum with $\alpha$=1 held fixed.
The dashed-dotted line is the H-atmosphere component, and the dashed
line is the power-law component.  The two spectral components are
equal near $\approx$3~keV, above which the power-law component
dominates, and below which the H atmosphere component dominates.  The
crosses are the observed \chandra\ data, with error-bars in countrate.
}
\end{figure}

\clearpage
\pagestyle{empty}
\begin{figure}[htb]
\PSbox{fig1.ps hoffset=-80 voffset=-80}{14.7cm}{21.5cm}
\FigNum{\ref{fig:chandraspec}}
\end{figure}
\newcommand{\LH}{2}
\newcommand{\SingleSpace}{
  \renewcommand{\LH}{0.90}
  \def\baselinestretch{\LH}
  \tiny
  \normalsize
}
\SingleSpace

\begin{deluxetable}{lr}
\tiny
\tablecaption{\label{tab:fits} \chandra\ Spectral Model Parameters (0.5-8 keV)}
\tablewidth{7cm}
\tablehead{
\colhead{Parameter} & 
\colhead{Value} \\
}
\startdata
\cutinhead{H Atmosphere }\\
\nhtt		& 0.30\ppm 0.06 	\\
\kteffinfty (eV) & 156\ppm18 \\
\rinfty\ (km)	 & 9.4\ud{2.7}{2.4}\\
Total Model Flux &  12		\\
\chisqrnu/dof (prob) & 1.7/10 (0.07)	\\
\cutinhead{H Atm.  + Power Law ($\alpha=1$)}\\
\nhtt		& 0.35\ud{0.08}{0.07}	\\
\kteffinfty (eV)& 135\ud{18}{12}\\
\rinfty\ (km)	& 13.4\ud{5}{4}\\
$\alpha$	& (1.0)			\\
$F_{X, PL}$	&  2.2\ppm1	\\
Total Model Flux &  14.7	\\
\chisqrnu/dof (prob) & 0.55/9 (0.84)	\\
\cutinhead{H Atm. + Power Law ($\alpha=2$)}\\
\nhtt		& 0.36\ud{0.10}{0.06}	\\
\kteffinfty (eV)& 131\ppm19		\\
\rinfty\ (km)	& 13.8\ppm4 \\
$\alpha$	& (2.0)			\\
$F_{X, PL}$	&  3.1\ud{1.6}{1.4}	\\
Total Model Flux &  14.9	\\
\chisqrnu/dof (prob) & 0.66/9 (0.72)	\\
\cutinhead{Photon Power Law}\\
\nhtt		& 0.72\ppm0.09 \\
$\alpha$  	& 4.1\ppm0.3 \\
Total Model Flux &  52\ud{13}{10}		\\
\chisqrnu/dof (prob) & 1.4/10 (0.16) \\
\cutinhead{Raymond-Smith}\\
\nhtt		& 0.47\ud{0.06}{0.05} \\
$Z$ (\zsun) & $<$5\tee{-3} \\
$kT$ (keV)	& 0.77\ppm0.10 \\
$\int n_e\, n_H dV$ \perval{cm}{-3}& (1.8\ppm0.4)\tee{57}\\
Total Model Flux &  19.4	\\
\chisqrnu/dof (prob) & 1.7/9 (0.08) \\
\cutinhead{Multicolor Disk}\\
\nhtt			& 0.36\ud{0.06}{0.05}\\
$T_{\rm in}$ (keV) 	& 0.43\ppm0.04 \\
$R_{\rm in}\sqrt{\cos(\theta)}$ km& 0.81\ud{0.23}{0.17}	\\
Total Model Flux	& 14.0		\\
\chisqrnu/dof (prob) & 1.74/10 (0.06)	\\
\cutinhead{Blackbody}\\
\nhtt			& 0.23\ppm0.06 \\
\kteffinfty\ (eV)  & 330\ppm20 \\
\rinfty\ (km)	 &  1.9\ppm0.3	\\
Total Model Flux & 10 		\\	
\chisqrnu/dof (prob) & 2.0/10 (0.03) \\
\enddata \tablecomments{X-ray fluxes are un-absorbed, in units of
\ee{-13} \cgsflux\ (0.5-10 keV).  Upper-limits and uncertainties are
90\% confidence. Values in parenthesis are held fixed.  Assumed source distance d=5 kpc.}
\end{deluxetable}

\begin{table}[htb]
\begin{center}
\tablewidth{40pt}
\caption{\label{tab:obs} Observation List}
\begin{tabular}{lclll} \tableline \tableline
Instrument	&  Obs.  Date	&  Days Since	& Days Since 	& Refs. \\
		& (dd/mm/year)	&  Outburst End	& Outburst Start&       \\	
		&		&		&		&       \\ \hline		
\rosat/PSPC (1)	& 15/10/1992	& 110-130	& 190		& 1     \\	
\rosat/PSPC (2)	& 24/03/1993	& ...		& 125		& 2, 3  \\	
\asca		& 21/10/1996	& 70		& 130		& 2, 4  \\	
\chandra	& 28/11/2000	& 7		& 70 		& 2     \\ \hline
\tablenotetext{}{1, \citenp{iauc5507,iauc5551}; 
2, RXTE/ASM; 
3, \citenp{iauc5664,iauc5665}; 
4, \citenp{iauc6416}
}
\end{tabular}
\end{center}
\end{table}

\begin{table}[htb]
\begin{tiny}
\begin{center}
\tablewidth{40pt}
\caption{\label{tab:results} Multi-Observation Spectral Fits}
\begin{tabular}{lccccccccccc} \tableline \tableline
			&  		&  \kteffinfty	& \rinfty\	&          & $F_X$	& $F_X$	       & Variable &	       &  \rosat  &    \rosat          & 			\\
\chisqrnu/dof (prob)	& \nhtt		&  (eV)		& (km)		& $\alpha$ & (PL)$^c$	& (Therm)$^d$  & Parameter& \chandra$^b$& PSPC (1)$^b$&PSPC (2)$^b$&\asca$^b$ 	\\	\hline
1.43/66 (0.01)		& --		& 150\ud{8}{11} & 9.8\ud{1.6}{1.4}& \nodata   & \nodata	& 11$^a$ 	&\nhtt	& 0.28\ppm0.04 & 0.47\ppm0.06 & 0.20\ppm0.04& 0.47\ppm0.06	\\ 
0.98/66 (0.52)		& 0.26\ppm0.03	& --		& 8.3\ud{0.6}{1.2}& \nodata    & \nodata&  11$^a$    & \kteffinfty& 164\ud{10}{6}& 145\ud{9}{5}&168\ud{12}{6}&153\ud{10}{4} \\
0.96/66 (0.58)		& 0.26\ppm0.03	& 159\ud{12}{7} & --		& \nodata   & \nodata	&  11$^a$     & \rinfty	& 8.7\ud{0.9}{1.5}&6.8\ud{0.7}{1.1}&9.3\ud{1.0}{1.6} & 7.5\ud{0.8}{1.3} \\
1.0/64 (0.49)		& --		& 120\ud{10}{30}& 17.6\ud{12.6}{3.7}&1.3\ud{1.3}{1.0}& 2.6 &  12     &	\nhtt	& 0.40\ppm0.08  & 0.60\ud{0.03}{0.09}&0.32\ppm0.07&0.60\ppm0.06	\\
0.62/64 (0.99)		& 0.31\ud{0.10}{0.04}&	--	& 12.0\ud{5.3}{2.3}& 0.80\ppm1.6     & 2.4 & 12$^a$  & \kteffinfty&143\ud{10}{30}& 127\ud{15}{21}& 148\ud{10}{30}&134\ud{8}{30}\\
0.60/64	(0.99)		& 0.31\ud{0.07}{0.02}& 141\ud{14}{23}&--	& 0.2\ud{1.6}{1.9}   & 2.5 & 12$^a$  &  \rinfty    & 11.6\ud{4.6}{2.3}& 9.1\ud{3.2}{1.8} & 12.5\ud{4.8}{2.2} & 10\ppm1.8 \\
0.71/64 (0.96)		& 0.55\ppm0.10	& 100\ppm30	& 16\ppm12	& 3.5\ppm0.5   	     &  -- & 5.3     & $F_{X,PL}$ & 2.4	& 1.2	&2.9	& 1.7		\\ \hline
\tablenotetext{}{Uncertainties are 1$\sigma$. Assumed source distance
d=5  kpc.  Model fluxes are corrected for absorption, in units of \ee{-13} \cgsflux. $^a$ Value is for \chandra\ observation. $^b$ Labels refer to observaitons listed in Table~\ref{tab:obs}. $^c$ Best-fit flux for the power-law component. $^d$ Best-fit Flux for the thermal component. 
}
\end{tabular}
\end{center}
\end{tiny}
\end{table}

\end{document}